\documentclass[electronic]{vgtc}           

\graphicspath{{figures/}{pictures/}{images/}{./}} 

\usepackage{times}

\usepackage{tabu}                     
\usepackage{booktabs}                  
\usepackage{lipsum}                   
\usepackage{mwe}                      

\usepackage{mathptmx}                  
\usepackage{subfiles}

\usepackage{enumitem}
\usepackage{float}

\onlineid{0}

\vgtccategory{Research}

\vgtcinsertpkg

\title{Sounds Uncertain: Exploring the Affective Aspects of Sonification for Uncertainty Visualization}

\author{Marcel-Simon Dutt\thanks{e-mail: st166580@stud.uni-stuttgart.de} 
\and Sita A. Vriend\thanks{e-mail: sita.vriend@visus.uni-stuttgart.de} 
\and Elias Elmquist\thanks{e-mail: elias.elmquist@visus.uni-stuttgart.de} 
\and Daniel Weiskopf\thanks{e-mail: daniel.weiskopf@visus.uni-stuttgart.de}}
\affiliation{\scriptsize University of Stuttgart}

\teaser{
  \centering
  \includegraphics[width=\linewidth]{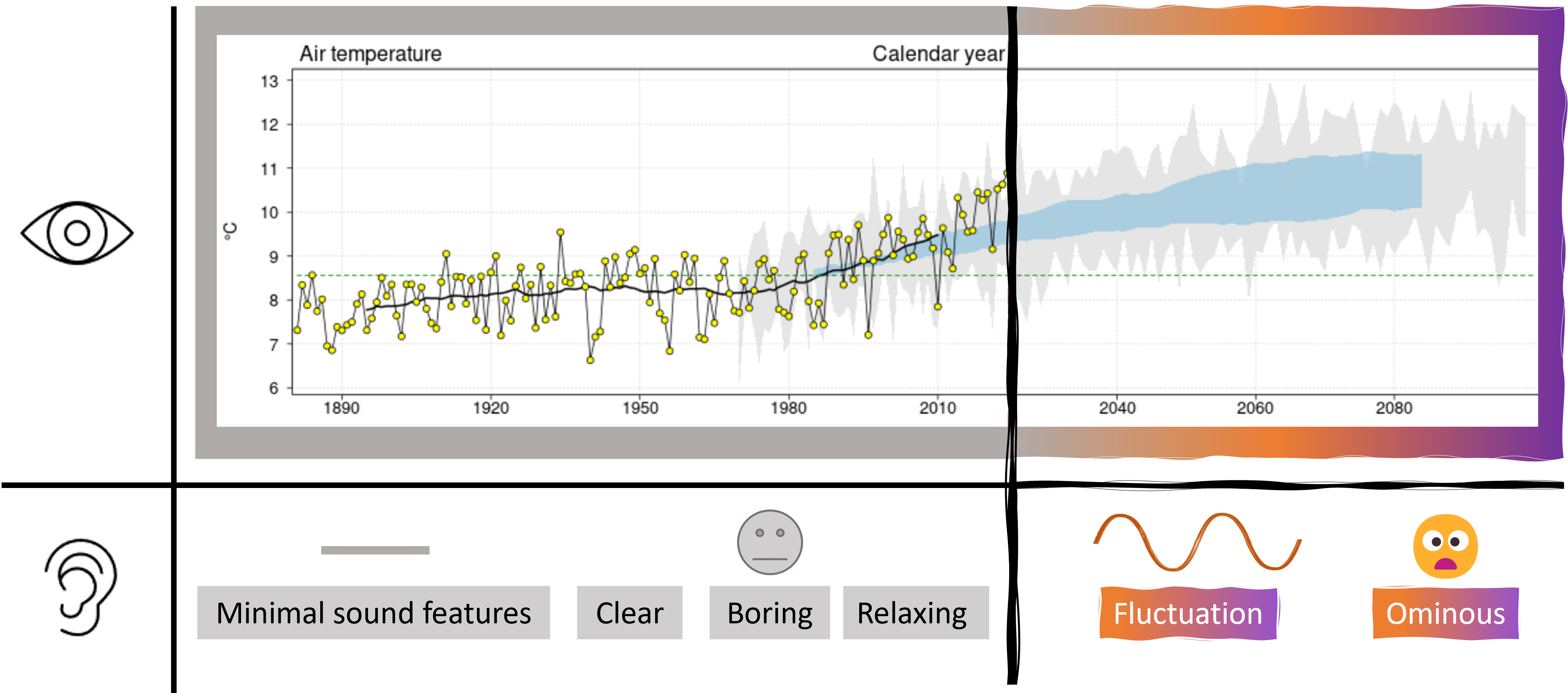}
  \caption{We explored what sonification qualities evoke the affective aspects of neutrality for the visualization of historic temperature data (seen on the left side of the image), and of uncertainty visualization for forecasted temperature data (seen on the right side of the image). Through a co-design user study, participants created a sonification that evoked the emotion of uncertainty, and a sonification that conveyed neutrality.
  }
  \label{fig:teaser}
}

\abstract{
Affective visualization can influence how users perceive, interpret, and engage with data by embedding and conveying emotion through visual design. 
While sound is widely used in media to evoke emotions, little is known about how sonification can support affective visualization. 
In this work, we investigate how sonification can communicate emotion in uncertainty visualizations through a co-design study. Participants created two sonifications to accompany a visualization: one conveying the affective component of uncertainty and one conveying neutrality. 
Our findings show that uncertainty was commonly associated with wavy auditory qualities related to an ominous sentiment. 
On the other hand, neutrality was associated with clear and relaxing auditory qualities.
These results provide insights for the design of visualizations that integrate sonification to communicate the affective component of uncertainty. 
}


\begin{document}

\firstsection{Introduction}

\maketitle
Affective visualization moves beyond representing data to embed the emotional context of the data~\cite{lan2024affective} or to utilize the benefits certain emotions provide with regards to the ability to retrieve and trust information, process stimuli, and influence decision‑making \cite{lerner2024emotions}.

Uncertainty visualization communicates data uncertainty such as error and probabilities~\cite{padilla2020uncertainty}, which can be difficult to understand~\cite{hoekstra2014robust}.
Especially lay people consider uncertainty a vague term~\cite{ballatore2019sonifying}.
However, uncertainty and its quantification has an affective component.
For example, not knowing the outcome of an important decision (e.g., due to a lack of information) can be nerve-wrecking and future predictions can be scary or hopeful. 
Such emotions are often easier to comprehend compared to complex constructs such as data uncertainty.
Affective visualization could therefore be suitable to improve the understanding of uncertainty in data.

Affective visualization has largely overlooked alternative modalities, including the auditory.
Sound is commonly used in synergy with visual media to communicate emotion to film and video-game audiences~\cite{lee2026emotion}. Players of the Super Mario Bros games might remember the change in soundtrack as they approached a boss-fight. The film Jaws employs the iconic two-note ostinato that represents the shark. In both examples, sound is used to build tension and induce fear.
The use of auditory channels to communicate data---sonification---is gaining traction in the scientific community \cite{Dubus2013AQuantities}, and has been integrated with visualizations \cite{enge2024open}.
Combining sonification and visualization can enable users to more easily grasp information and gain new insights into the represented information \cite{enge2024open, hermann2011sonification}, including uncertainty \cite{vriend2025semiotics}.
Affective aspects of sonification can be utilized to influence how data is perceived and understood, to allow researchers and designers reduce bias, and to \mbox{foster engagement~\cite{middleton2023data}.}

This paper investigates sonification qualities that evoke the affective aspects of uncertainty and how these can enrich uncertainty visualizations.
We additionally examine sonification qualities perceived as neutral. This serves as a baseline for comparing auditory qualities and reasoning intended to convey uncertainty, and provide insights into how designers can avoid unintentionally eliciting emotional responses.
The following research questions were formulated to reflect these aims:

\textbf{RQ1}. What sonification qualities induce the affective aspects of uncertainty?

\textbf{RQ2}. What are the characteristics of a sonification that conveys neutrality? 

For this purpose, a user study employing a co‑design approach was conducted. Participants were tasked with creating sonifications that matched a visualization of future, predicted temperatures that should feel uncertain, and historic temperatures that should feel neutral. 
This paper is based on, and largely extends, initial thesis work by one of the authors \cite{dutt2025co} supervised by the co-authors. 
The study was preregistered~\cite{dutt_vriend_elmquist_2025}.

Our research offers the following contributions to the uncertainty visualization community:
\begin{enumerate}
    \item Design insights into how to integrate sonification in visualization to convey the affective aspects of uncertainty \mbox{and neutrality.}
    \item Insights into the reasoning behind sonification design decisions users made to convey uncertainty and neutrality.
    \item The neutral and uncertainty sonifications created by each participant as supplemental material \cite{DARUS}.
\end{enumerate}

\section{Related Work}
To understand how the auditory modality can complement the visualization of uncertainty, we examine related work in affective visualization, uncertainty visualization, and sonification to provide a background and identify research gaps.

\subsection{Affective Visualization}
Emotions such as fear, sadness, and desire influence how visual stimuli are perceived and cognitively processed~\cite{zadra2011emotion}. 
Building on this understanding, affective visualization explores how design choices can evoke emotional responses in users and aims to embed emotion in visualization, or express emotion through visual design.

Visual channels such as color and motion~\cite{feng2017beyond, Lan2022Kineticharts, Martin2025health} and the use of specific chart types~\cite{Blair2025emotional} can evoke emotions.
By altering users' emotional states, affective visualization can shape how people interpret, evaluate, and act upon visualized data, affecting trust, information recall, comprehension, engagement, and visual judgments, all of which can significantly influence decision-making~\cite{Harrison2012emotion, lerner2024emotions}.

In contrast, visualization can express emotion to convey context and meaning of the underlying data, adding an extra dimension of the data to the visualization.
Color palettes can communicate the meaning and context of data because colors are often associated with specific emotions. 
For instance, blue is often associated with sadness, which is reflected in language when someone is ``feeling blue.''
Consequently, selecting an appropriate palette is important for emotionally charged data. For example, bright, cheerful colors are inappropriate for visualizing murder cases, as they contradict the mournful context of the data \cite{Anderson2022congruence, Bartram2014color,Braun2026color,kushkin2023cognitively}.

Affective visualization is a fundamental component in storytelling approaches to visualization~\cite{lan2024affective, lan2022emotions}. For example, Kineticharts~\cite{Lan2022Kineticharts} convey emotion and expressiveness and improve engagement without hindering data comprehension.
However, affective visualization has largely overlooked alternative perceptual modalities such as the auditory, and their role in uncertainty visualization remains unexplored. Therefore, we investigate the use of sonification to communicate the affective aspects of uncertainty \mbox{in visualizations.}

\subsection{Uncertainty Visualization}
Uncertainty visualization encompasses many aspects of uncertainty~\cite{ padilla2020uncertainty} that can happen in any stage of the visualization process \cite{Hagele2022uncertainty}:
from precision and accuracy of the collected data \cite{MacEachren01012005}, to missing data or a lack of information \cite{Skeels2008uncertainty}, to risk or probabilities~\cite{spiegelhalter2017risk}, to uncertainty data resulting from predictions.
Uncertainty is often perceived as vague, especially by lay users, making it hard to understand~\cite{ballatore2019sonifying,hoekstra2014robust}.
While alternative modalities such as sonification have been used in combination with visualization to improve understanding \cite{vriend2025semiotics}, uncertainty remains difficult to grasp.

However, uncertainty also has an affective dimension that is often experienced negatively and associated with feelings of anxiety \cite{gray1982neuropsychology,grupe2013uncertainty,hirsh2012psychological}.
Not knowing due to missing information can evoke feelings of anxiety and risk has a scary connotation. On the other hand, uncertainty can also be linked to the concept of hope~\cite{vazard2024feeling}.

The affective aspect of uncertainty has not yet been utilized in uncertainty visualization. Incorporating affect could enrich visualizations by communicating both data uncertainty and its emotional meaning and context.

\subsection{Sonification}
Like visualization, sonification aims to convey information. While visualizations use visual channels such as color, size, and shape, sonifications employ auditory channels such as pitch and loudness. 
The modalities can be combined to utilize additional perceptual channels or to emphasize already visualized data \cite{enge2024open}. 

One such approach is to use sonification to convey affective aspects of data, as sound and music can convey and evoke emotions~\cite{picard_affective_1997}. 
In addition, emotional cues from sound have been found to influence visual perception~\cite{tajadura-jimenez_auditory-induced_2008}. 
Hence, sonification could be used to enhance affective visualizations. Prior work explored this idea through data melodification for affective data experiences~\cite{zhang2025data} and emotional musification frameworks based on arousal and \mbox{valence~\cite{Godbout2018Musification}.} 

Previous research demonstrated benefits of integrating sonification into visualization for conveying affect.
For example, Rönnberg~\cite{ronnberg2021emotion} found that sonification was effective for conveying emotions in a visualization of running and weather data. 
In another study, Nath~\cite{nath2020hearfeardatasonification} added an auditory icon of a woman screaming to emphasize the traumatic aspects in a visualization of crime against women, which was considered impactful by participants.

Vriend et al.~\cite{vriend2025semiotics} examined users' audiovisual mapping preferences for uncertainty based on auditory and visual channels identified in earlier research on semiotics of uncertainty \cite{ballatore2019sonifying, maceachren2012visual}. In this paper, we instead investigate which auditory parameters users associate with an affect of uncertainty and neutrality and why. 
We included modulation and white noise beyond the auditory channels studied by Vriend et al.~\cite{vriend2025semiotics} (loudness, tempo, and pitch). The limitation to a few channels avoided overloading participants with choices and was complemented by our co-design methodology, which allowed participants to talk about auditory parameters unavailable to them in the study tool. 

Participants' reasoning for the mappings remained unclear in these prior works. It is likely that participants associated auditory channels with the emotional associations of uncertainty.
We therefore examine the relation between sonification design reasoning and affective aspects of uncertainty more closely in order to understand how sonification can provide the emotional meaning and context of uncertainty visualization.

\begin{figure*}[!ht]
    \centering    \includegraphics[width=1.0\linewidth]{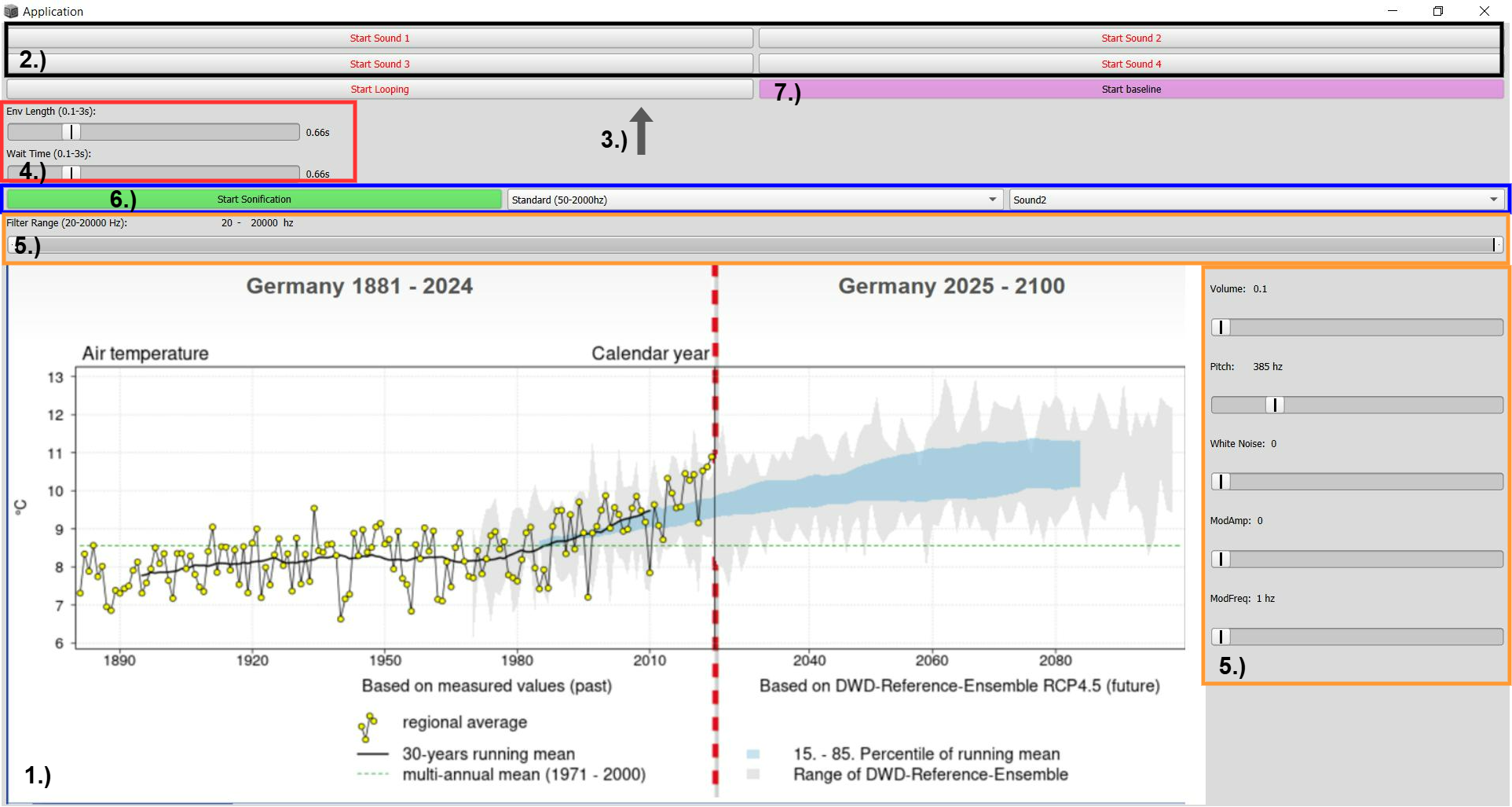}
    \caption{The study tool consists of seven main parts: \textbf{1.)} The visualization showing the yearly average German temperatures since 1881, including predicted temperatures up to 2100.
    \textbf{2.)} Four white buttons for selecting a musical instrument (labeled as Sound 1-4 to reduce bias). \textbf{3.)} One looping button for looping the sonification. \textbf{4.)}~Sliders to change the envelope length and wait time of the looping sonification. \textbf{5.)} Sliders that allow changing the auditory parameters for the currently selected musical instrument. \textbf{6.)} A green start button and drop-down menus allowing participants to map the pitch to the given data on the left side. The middle drop-down menu lets participants choose their pitch range based on pre-set options. The right drop-down menu allows for the change of the instrument, while the green button plays a sonification based on the options selected in the blue, orange, and red boxes. \textbf{7.)} A pink button to play a pre-definded sonification, also called the baseline. The visualization is publicly available from Deutscher Wetterdienst (\url{https://www.dwd.de/DE/klimaumwelt/klimaatlas/klimaatlas_node.html}).}
    \label{fig:GUI}
\end{figure*}

\section{Study Method}
We conducted a user study with a co-design approach to investigate sonification qualities that induce affective aspects of uncertainty and neutrality.
Co‑design approaches encompass methods aimed at democratizing the design process and often involve inviting users to create the artifacts of which they are the target audience~\cite{Steen2013codesign,Visser01042005}.
Although users are not design experts, the artifacts they design and the decision‑making processes behind them offer valuable insights \cite{koch2026multimodal}. 
These approaches are gaining traction in visualization  \cite{Goodwin2013energy, janicke2020participatory,Khowaja2022Participatory, Musaeus2026Participatory} and sonification research \cite{droumeva2006role,grond2019participatory}.

Prior sonification research showed that auditory mappings chosen by designers do not always align with user interpretations \cite{Walker2005mappings}. 
To address this gap, we employed a co-design approach and invited participants to design their own sonifications. 
They configured auditory channels to express uncertainty and neutrality in a visualization of past and future temperatures. 
The resulting sonification artifacts, together with participants’ reasoning, provide insights for designers seeking to convey affective dimensions of uncertainty through sonification in uncertainty visualizations.

The study was approved by the ethics commission of the University of Stuttgart. Written consent was given by the participants. The study was preregistered on the Open Science Framework \cite{dutt_vriend_elmquist_2025}.

\begin{figure*}
    \centering
    \includegraphics[width=1.0\linewidth]{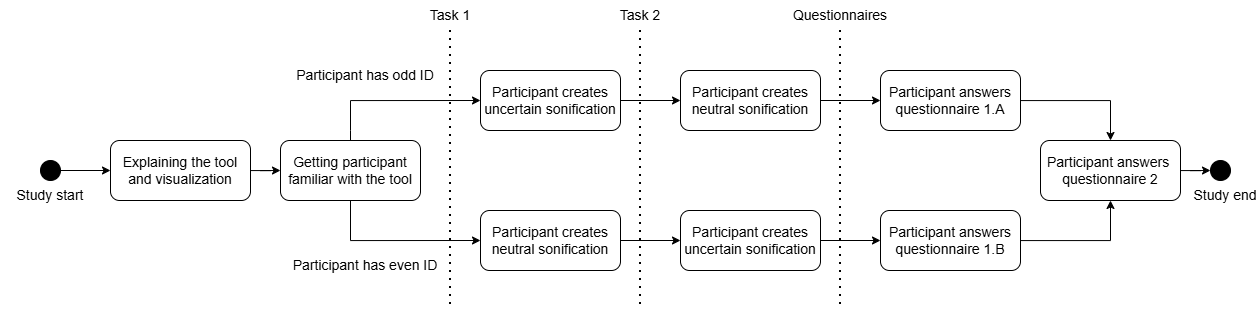}
    \caption{The study procedure started with a tutorial of the tool, and continued with the creation of an uncertain and neutral sonification. This was followed up with filling in two questionnaires.}
    \label{fig:Study procedure}
\end{figure*}

\subsection{Study Tool}
We created a tool, as seen in~\autoref{fig:GUI}, to allow non-expert participants to create their own sonification using different buttons, a set of sliders, and drop-down menus. 
Only a limited number of parameters were available to avoid overwhelming participants with excessive choices. Additionally, the tool acted as a conversation starter during the co-design process. Although participants may not have been able to create sonifications that exactly matched their intended designs, the tool enabled them to express, discuss, and reflect on their ideas.
Hence, we chose auditory parameters that 1)~were intuitive to understand, 2) made a noticeable difference when changed, 3) worked together to offer participants a wide range of options, and 4) were distinct from one another.
The visualization used in the study depicted historical temperatures in Germany from 1881 to 2024 and future predicted temperatures from 2025 to 2100. 
Historical temperatures represent recorded observations and therefore do not contain predictive uncertainty, thus could be considered neutral. 
In contrast, future temperatures are predictions visualized using uncertainty bands to communicate the uncertainty associated with these forecasts.

\paragraph{Instrument:}
Four buttons allowed the user to select an instrument with a different timbre, or tone quality.
The instrument choices were grounded in previous work that states that each musical instrument has a different emotional affect on a listener to varying degrees, due to their timbre, even if the instruments a part of the same family such as trumpets and trombones \cite{eerola2012timbre, wu2013spectral}. The following instruments were included in the tool: a piano, a deep brass instrument such as the horn or tuba, a light brass instrument such as the trumpet, and a simple electric synth.  To minimize bias regarding the instruments, they were labeled with non-telling names in the tool: Sound 1 to Sound 4, respectively.

\paragraph{Volume:}
The volume slider controls the overall loudness of the auditory output. 
Loudness is considered to be impractical for sonification~\cite{neuhoff2002pitch, hermann2011sonification}. However, it was included since loudness has been shown to be an intuitive mapping for uncertainty~\cite{ballatore2019sonifying,vriend2025semiotics}.

\paragraph{Pitch:}
The pitch slider controls the fundamental frequency, which is either applied to a selected instrument or is added as a specific frequency to the active sonification~\cite{Kollmeier2008}. Pitch was included because it is the most frequently used auditory parameter in sonification mappings~\cite{Dubus2013AQuantities}. 

\paragraph{White noise:}
White noise can be added to the individual instruments of the sonification, with varying levels of loudness controlled with a slider. Continuous white noise has been found to be emotionally important but still categorized as negative~\cite{MASUDA201899}.
However, white noise in lower volume levels has also showed an increase in cognitive performance such as an increased attention span and accuracy, while also lowering stress levels~\cite{awada2022cognitive}.

\paragraph{Amplitude modulation:}
Amplitude modulation oscillates the volume of the sonification over time, creating a so-called tremolo effect. As it is a more specific auditory parameter, there is less related work regarding it. However, as it is continuously modulating the sound over time, it might induce some kind of emotion.

\paragraph{Frequency of amplitude modulation:}
This auditory parameter complements the amplitude modulation, by specifying how quickly the modulation should happen over time. At lower values, the oscillations produce a gentle vibrato, whereas higher values create rapid oscillations and additional overtones. 

\paragraph{Looping:}
Continuous auditory stimuli, such as music, have an effect on the emotional state of listeners, with especially tempo having an impact on the emotions of listeners \cite{YangMusicTempi2025}.
However, sonifications often use short bursts of sound rather than continuous tones, because brief sounds are easier to map to data and are more clearly perceived by listeners \cite{hermann2008taxonomy}.
The looping button changes the selected sound from a monotone, sustained tone to shorter, periodical notes. It is applied to the currently selected instrument and two sliders are used below it: The envelope length slider determines the length of the notes, and the wait time slider sets the time between notes. Both sliders cover a temporal range from 0.1 to 3 seconds. Enabling the looping also causes the tones of the selected instrument to decrease their pitch over the envelope length. 

\paragraph{Frequency filter:}
The frequency filter is controlled by a two-sided slider that allows participants to shape the spectral content of their sonifications. This applies to all aspects of the sound, including the added white noise.

\paragraph{Baseline button:}
Pressing the baseline button plays a standardized sonification to give a practical example of what a resulting sonification could sound like. The sonification maps the temperature data of each year to the pitch of a distinct note, which is a common mapping~\cite{Walker2002MagnitudeSonification}. 

\paragraph{Sonification button:}
The sonification button works the same way as the baseline button, but allows the user to change the sonification through the previously mentioned parameters. In addition to the pitch slider, the user can select frequency ranges that the sonification should use, varying from a wide range (50--2000\,Hz) to a more narrow range (800--1600\,Hz).

\subsection{Study Procedure}

\noindent 
The study took place in a quiet room and followed the procedure illustrated in \Cref{fig:Study procedure}.
After signing the consent form, participants were given a pair of Beyerdynamic DT 770 Pro headphones, and some time to get familiar with the study tool.
During the study, participants were tasked to create a 1) neutral-feeling sonification for the historic temperatures (left side of the visualization in \Cref{fig:GUI}) and 2) an uncertain-feeling sonification for the future predicted temperatures (right side of the visualization in \Cref{fig:GUI}) using the study tool.
We counter-balanced the sonification participants started with, to account for possible carry-over effects of perspectives from the first sonification (as shown in \Cref{fig:Study procedure}).

We employed a talk-aloud procedure to capture participants' understanding of uncertainty, neutrality, and reasoning behind their chosen auditory parameters while they created the sonifications.
Additionally, we captured voice recordings, screen recordings of interactions with the tool, and PC audio.
After creating their sonifications, the participant filled in one questionnaire regarding the participant's perspective of the study tool, and after that they filled in a demographics questionnaire.

\subsection{Participants}
\noindent 
Participants were recruited at the university campus via posters and word-of-mouth.
We recruited participants who were not sonification experts or professional musicians to address the interpretation gap between sonification designers and users \cite{Walker2005mappings}.

The study was initially conducted with 10 participants during the preliminary thesis work~\cite{dutt2025co} (1 woman, 9 men), resulting in insights largely shaped by men’s perspective. To reduce this gender imbalance and avoid bias in the patterns we observed, we then recruited 11 additional participants. 
The participant criteria and recruitment strategy remained the same.

\begin{figure}[t]
     \centering
     \includegraphics[width=0.7\linewidth]{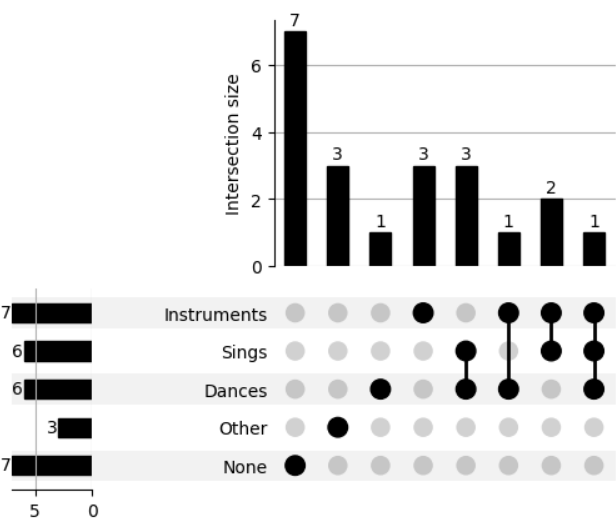}
     \caption{The distribution of musical inclination ratings of the 21 participants, which was recorded in a demographics questionnaire.}
     \label{fig:Musical inclination}
\end{figure}

A total of 21 participants took part in the study (7 women, 14 men).
All participants reported having normal or corrected-to-normal vision and hearing, reported not having a condition that affects sensory perception such as synesthesia. 
The mean age was 26.5 years, ranging from ages 23 to 32.
Out of these 21 participants, ten were students, three were PhD students, four were university researchers, and the remaining four participants were employed in occupations unrelated to music or sonification.
Eleven participants reported engaging in music-related hobbies, with more details shown in~\autoref{fig:Musical inclination}.
Only one participant reported prior familiarity with sonification, whereas 19 participants had either never encountered the concept or recognized it solely by name.

\section{Analysis and Results}
We collected screen and audio recordings of participants' interactions with the study tool, voice recordings of the talk-aloud protocol, and the created sonifications.
The transcripts of the talk-aloud protocol were analyzed according to Braun and Clarke’s reflexive thematic analysis (RTA) \cite{Braun01012006}.
RTA was used for its systematic identification and interpretation of found themes in the data, while still being flexible.
Initial codes and themes were generated based on transcripts of the first 10 participants during the first author's thesis~\cite{dutt2025co} under supervision by the co-authors.
The first author familiarized themselves with the data by reading transcripts several times and also viewing the study recordings. 
A top-down approach was used for the analysis of the transcripts.
Transcripts were carefully read while rough notes were taken in a separate file.
First, the participants' definitions of neutrality or uncertainty were established, then how they related their definitions to some known phenomenon, and finally how they tried to create sonifications through the auditory parameters.
After coding, each sonification was grouped into broader emerging themes, such as happy, sad, stressful, etc.
The themes were based on the overall qualities of the sonification and the given context by participants. 
The thesis identified three distinct themes: The simple theme, the wavy theme, and the ominous theme.
During the thesis work, neither of the co-authors independently coded any of the first ten transcripts, they however, offered insights into the procedure and helped with the refinement of codes and themes.

The transcripts of the 11 additional participants were first coded by the first author according to the codes and themes generated during the thesis.
The same top-down approach was used for the analysis of the additional transcripts.
To add an additional perspective, the second author went over all transcripts to supplement initial coding, thereby generating new codes, as well as merging and splitting original codes.
Throughout the analysis process, the first three authors regularly met to discuss suitability of codes assigned to quotes. They generated new codes, merged and split codes until an agreement was reached.
Themes were discussed in a similar manner, and a new theme was generated that was not present in the thesis.
Throughout the entire analysis process, the main focus was on which and why certain auditory parameters were chosen for each of the sonifications.

\begin{figure}[!ht]
    \centering
    \includegraphics[width=1.0\linewidth]{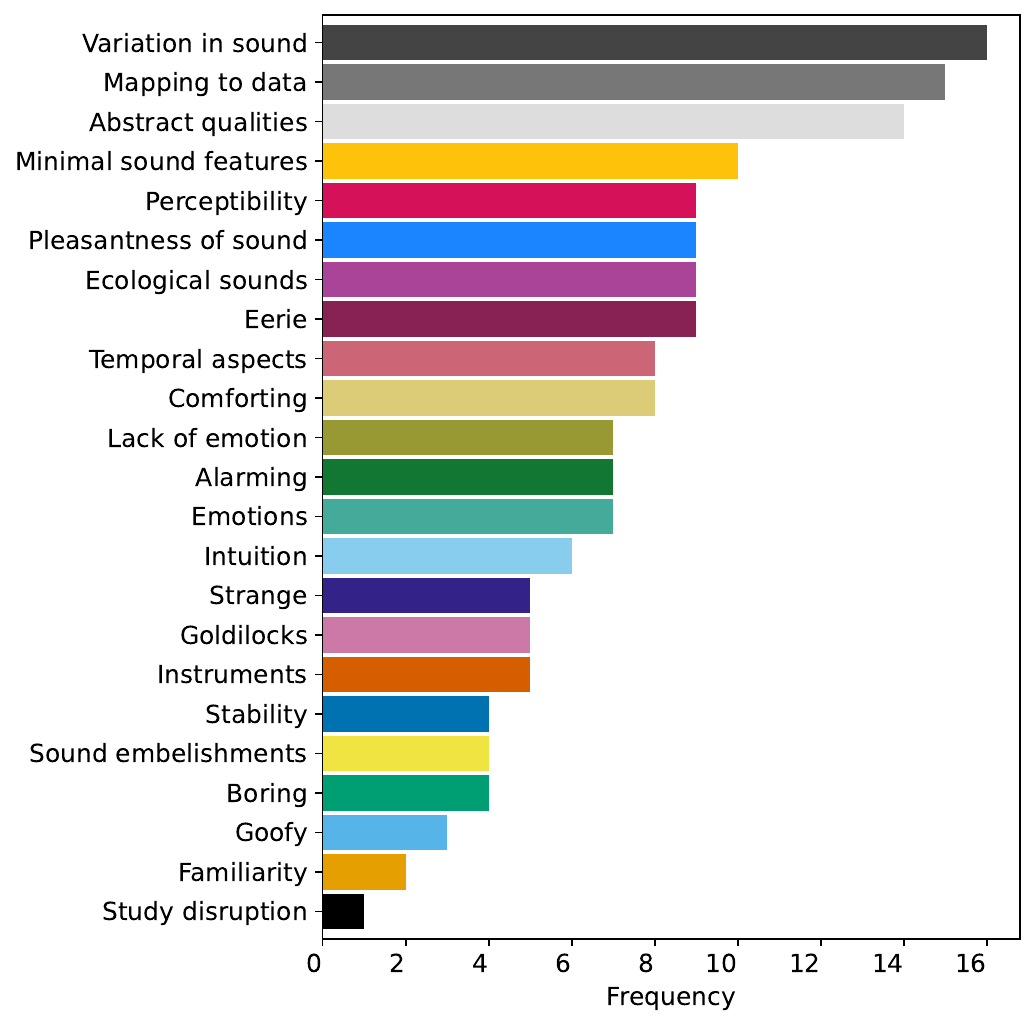}
    \vspace*{-4ex}
    \caption{Number of participants mentioning the code.} 
    \label{fig:Frequency_Codes_Unique-1}
\end{figure}

\subsection{Codes}
In this section, we present a description of each code identified in the RTA process and created a codebook resulting from this process~\cite{DARUS}. \Cref{fig:Frequency_Codes_Unique-1} shows the number of participants that mentioned each code. 

\subsubsection*{Ecological sounds}
Participants associated sounds with ecological phenomena, often related to memories \textbf{associated with emotions}. Consequently, these were avoided in neutral sonifications.
Participant P09 related to a personal experience with video games: ``\textit{This is just reminds me of when you play Mario and then die. And this is a bit too deep in my opinion. [...] it has something to it that doesn't make it neutral}.''
Some instruments, such as the piano, were more generally associated annoyance: ``\textit{But if this feels more like if the radio connection is gone, so you have to [reconfigure]. So for me, it brings annoyance, [which makes this sound] not neutral}''~(P03).

\subsubsection*{Instruments}
Some instrument options reminded participants of musical instruments, which made them unfit for both neutral and uncertainty sonifications.
When creating a neutral sonification, P15 stated: ``\textit{So, I feel like [the horn] and the [electric synth] [...] are sort of representing some sort of instrument to me. So, I'm not going to use them.}''
When creating the uncertainty sonification, P09 shared: ``\textit{I don't know if that's uncertain to me. It reminds me of some kind of guitar string, with a high frequency, but I think this is not uncertain.}''

\subsubsection*{Abstract qualities}
Contrary to \textbf{ecological sounds}, some participants related sounds to abstract phenomena when creating neutral or uncertain sonifications.
For example, P08 considered certain instruments neutral: ``\textit{[...] this [piano] sound and these two sounds [of the piano and trumpet], they feel more more alive more more light.}''
On the other hand, P19 says: ``\textit{Yeah because it sounds less certain [and] more like mysterious,}'' which made it more uncertain for them.

\subsubsection*{Emotions}
Some participants explicitly related to emotions when reasoning for and against using an auditory parameter.
High pitch was associated with positive emotions and low pitch with negative emotions. Hence, P08 used a pitch that was neither high nor low for the neutral sonification: ``\textit{[...] the higher notes give more of a like lively emotion, more of like a light emotion. The base sounds, the really dark bass sounds, they sound dark, they sound like melancholy.}''
Certain instruments and white noise were associated with emotions.
In combination with specific instruments, white noise was found suitable for neutral sonifications by P12: ``\textit{[...] [The electric synth] on its own has a very neutral emotion according to me. And white noise is just adding a little hint of an after effect of an emotion.}''

\subsubsection*{Eerie}
Uncertainty was considered eerie and often described as scary, uneasy, or ominous. 
Participants, including P10, related eeriness to \textbf{ecological sounds} such as scary movies:
``\textit{So, the first one sounds a bit, like, creepy. Like a thriller or a horror movie sound.}'' 
Especially low-pitched sound was considered fitting for uncertainty sonifications:
``\textit{I just correlate uncertainty with a kind of creepiness and that correlates also for me with a low tone}'' (P05).
White noise was additionally associated with uncertainty:
``\textit{White noise I found also disturbing and not neutral}'' (P06).

\subsubsection*{Strange}
Some auditory parameters, including pitch and some choices of instruments, were avoided for neutral sonifications, as they were considered strange:
``\textit{[...] if [pitch]'s too high, then it starts to sound a bit kind of strange. And it's almost like there are different emotions that kind of come through of, like, curiosity or discomfort}'' (P21).
Strange auditory qualities were not considered suitable for uncertainty sonifications either: 
``\textit{If I had to say what the like modulation makes me think of, it would be something else, maybe adjacent to uncertainty. [...] if you had data where you had outliers or something, then I might think to use something like that, because it gives a bit of a strange feeling. And then you might think, oh, there's something strange or unexpected going on here. Whereas uncertainty isn't really the same as seeing something unexpected}'' (P21).

\subsubsection*{Alarming}
Some auditory parameters such as volume and choices of instruments were considered alarming and therefore suitable for the uncertain sonification.
Regarding volume, P03 said: ``\textit{So in the beginning the volume would be lower and then over the years it would steadily increase. So that you get the feeling that something is wrong, gets bigger.}''
Alarming auditory qualities, such as high pitch were avoided for neutral sonifications. For instance, P05 stated:
``\textit{it would be too stressful, [if pitch] is too high,}'' which they ``\textit{would correlate it with an alarm or something like that.}''

\subsubsection*{Boring}
Participants mentioned that both neutral and uncertainty sonifications should sound boring.
For their neutral sonification, P08 said: ``\textit{Yeah so for me it's the most boring sound and I feel like that gives me the most neutral vibe right.}'' 
Contrary, P21 considered boring fitting their uncertainty sonification: ``\textit{whereas I wanted that kind of dull, like drawn out feeling for the uncertainty.}''

\subsubsection*{Goofy}
White noise, amplitude modulation, and
frequency of amplitude modulation
were considered funny, not serious, comedic, or goofy.
Hence, P05 avoided these qualities in the neutral sonification: ``\textit{Also white noise, mod amp, and mod frequency just sound goofy. I just have to laugh when I use them so it's not neutral anymore if I use them.}''

\subsubsection*{Comforting}
Certain instruments and even white noise were considered appropriate for a neutral sonification since they were considered comfortable.
Multiple participants, including P16, considered white noise comforting, thus fitting a neutral sonification: ``\textit{Yeah white noise adds a background humming sound which is a more natural sound, like a raining environment. There is some [humming] sound.}''

\subsubsection*{Intuition}
Participants often initially stated that they did not know why they had selected particular auditory parameters, describing their decisions as something that ``felt right.'' Ultimately, most participants articulated reasons for their choices.
However, in some cases participants were unable to reason beyond intuition or gut feeling:
``\textit{It's hard to say because of what I'm feeling is just like a gut feeling}'' (P08).

\subsubsection*{Familiarity}
Some participants used an instrument because they heard it earlier during the study session.
For instance, when creating the uncertainty sonification, P08 said: ``\textit{[...] with sound four it kind of stuck in my head.}''

\subsubsection*{Goldilocks}
Goldilocks is a fairytale character who finds she prefers porridge that is neither too hot nor too cold.
Like the character, participants wanted auditory qualities that were ``just right'' for their neutral sonification. The pitch needed to be ``just right'' for P05: ``\textit{[...] the pitch slider for me is in the middle of it, because I just wanted a neutral tone.}''

\subsubsection*{Lack of emotion}
Instruments that did not evoke emotion were viewed as appropriate for neutral sonifications: ``\textit{All the other sounds had a happy sensation to it [...] but sound three was uh I didn't dislike it nor enjoy it which was just neutral}''~(P13). 
White noise was avoided by P11 in their neutral sonification: ``\textit{[...] when I'm increasing the white noise, it just gives uncertain vibes. So, I think I'll keep it less. Like this is not inducing any emotion, not uncertainty or happy or anything.}''

\subsubsection*{Mapping to data}
Especially for neutral sonification, participants wanted to map auditory parameters to the visualized data. Both the data aspects and the auditory parameters chosen varied per participant.
Often, participants wanted to sonify the regional average:
``\textit{I tried to sonify those [yellow] dots}'' (P09). 
Contrary, participants like P06 chose to map to the 30-year running mean since mapping to the regional average would introduce too much auditory variation:
``\textit{Because the oscillations between two [following] years could be big, but overall you see the mean isn't really changing that much. [...] So I would not represent the singular years but the mean overall and in this case this looped sound with maybe a higher pitch at the end would represent it very well.}'' 
Mapping to data was also considered for uncertainty sonifications. Some participants wanted to sonify the whole range of the uncertainty bands, whereas others like P07 mapped the the outermost uncertainty band on the prediction side: ``\textit{And then I tried to mimic by using the pitch, the uppermost boundary of the gray area. So if it's higher, I made the pitch higher. When it went lower, I made it lower.}'' 

\subsubsection*{Minimal sound features}
Especially neutral sonifications were deemed to sound clean and contain minimal amount of different sound features.
Participants such as P19 relied on pitch and volume alone: ``\textit{I want to have a clear sound and yeah I just kept the pitch and volume.}'' 
Wanting minimal features is closely related to \textbf{mapping to data}
as participants like P09 wanted to keep the focus on the data: ``\textit{I just try keep the background noise away to keep it more neutral. And tried to just have these dot like sounds.}''

\subsubsection*{Sound embellishments}
Opposed to \textbf{minimal sound features}, some participants wanted to add sound features to their sonifications.
White noise was added to uncertainty sonifications by P20:
``\textit{If [the sonification is] clear, then one knows [the data], with no white noise. For me, the noise adds uncertainty to it.}''
Contrary, P08 added white noise to their neutral sonification:
``\textit{I wish could I could put [...] a continuous sound of white noise over it and not like it with every note playing or without with every sound playing every data it restarts the white noise.}''

\subsubsection*{Perceptibility}
Sonifications should be audible and allow listeners to perceive the information, often with adequate volume and by removing unnecessary auditory elements. For neutral sonifications, participant such as P02 avoided sound designs in which differences between data values were difficult to distinguish: ``\textit{I find it easier to follow the data, because I have the direct comparison, just like with pictures, I guess. If you see them rapidly and interchange, you see differences, and if it's very close to each other, I can much easier detect that it's going up or down, and how it compares to the ones next to it. So, that's why I wanted it quick.}''

\subsubsection*{Pleasantness of sound}
Participants thought that a sonification, both uncertain and neutral, should sound pleasant. Certain instruments were used because they were considered pleasant, whereas others were avoided because they were seen as unpleasant. For example, P07 avoided sound 1:  ``\textit{[...] it's kind of pissing me off to hear the sound for more than five seconds.}''
Unpleasant sounds were also avoided because they were associated with \textbf{emotion}: ``\textit{[...] if you decrease the noise, there's some irritating sound. So it might, give you some emotion. That's why I increased it}'' (P14).

\subsubsection*{Temporal aspects}
Temporal aspects such as wait time were controlled to slow down or speed up. For example, P11 increased wait time because they related slowness to uncertainty: ``\textit{[...] for some reason I feel they are, like, uncertain in terms of like, slow uncertainty}.'' 
However, while short wait time was considered neutral by some participants, others, including P21, related longer wait time to neutrality: ``\textit{[...] given that I want the sound to be neutral, it felt like if you want to convey data well.}''

\subsubsection*{Variation in sound}
Variation in sound included wanting randomness and oscillation, especially for the uncertainty sonification. Participants used a variety of auditory parameters including frequency modulation and pitch variation.
White noise was used by P21 to represent all possible future temperatures: ``\textit{Like I added in this white noise because it makes the sound feel a bit more like it has like more depth to it or like more variation. So it kind of represents a wider area rather than like a specific point.}''

\subsubsection*{Stability}
Opposed to \textbf{variation in sound}, participants expressed that neutral sonification should sound stable.
Auditory parameters that introduced fluctuation were avoided, even when participants, like P03 wanted to \textbf{map to data}:
``\textit{So the air temperature is rising. So this also something that is changing. [That] can't be neutral, in my opinion. [...] So change can be either good or bad, but not neutral}.''

\subsection{Themes}
Themes capture recurring patterns in participants’ statements. The codes were clustered and labeled to reflect their underlying meaning. Four themes were generated: \textbf{1.)} the \textbf{clear theme}, \textbf{2.)} the \textbf{relaxing theme}, \textbf{3.)} the \textbf{wavy theme}, and \textbf{4.)} the \textbf{ominous theme}, each linked to either neutrality or uncertainty and described by consistent codes and auditory parameters. The sonifications were sorted into the four themes, each corresponding to the specific task each participant completed. Therefore, a sonification created for a neutral task was never placed into an uncertain theme, regardless of the codes, the used auditory parameters, or the participants’ explanations. 
Although some participants omitted the typical code, their expressed sentiments and comparable auditory parameters matched those of the theme’s members, justifying their inclusion.

\subsubsection*{Clear}
The \textbf{clear theme} can be used to describe the neutral sonifications and includes the codes \textbf{minimal sound features} and \textbf{lack of emotions}. It is described through the absence of more complex sound features, such as white noise or modulation. 
Participants mainly focused on having a clear sound quality that would not evoke any kind of emotion.
This led the instrument choices to be monotonous, as hearing a musical instrument could lead to some emotional connection to the sound, which then would not feel neutral anymore.
As such, the electric synth was by far the most used instrument for this theme. 
Volume did not play a role in this theme, with most participants stating they just set it to a personally pleasant level.
\textbf{Mapping to data} was prevalent as most participants chose to map pitch directly to the visualization through the use of the sonification button.
However, there was a small subset of participants who decided not to do so.
A few participants also shared the sentiment that a neutral sonification should sound \textbf{boring}, with some participants connecting this feeling to personal experiences, such as the beeping sound of a phone when it is trying to connect.
This was conveyed through an increase in wait time.
However, the differences in used auditory parameters were not significant to be considered their own theme.

\subsubsection*{Relaxing}
The \textbf{relaxing theme} is a considerably smaller theme that can also be used to describe neutral sonifications.
Participants did not equate neutrality with an absence of emotions but rather with feeling \textbf{comfort}. 
Their differences stemmed mainly from the reasoning behind their choice of auditory parameters, especially the use of white noise and modulation. By contrast, the choices were grounded in emotion, with participants creating sounds they themselves considered comforting.
Participants also expressed that they wanted background noise, as this would make the experience less silent and more comforting.
One participant mentioned how white noise adds a natural, humming background reminiscent of a raining environment, an ecological sound that they associate with comfort and, \mbox{thus, neutrality.}

\subsubsection*{Wavy}
The \textbf{wavy theme} can be used to describe the uncertain sonifications.
The most common code found in this theme is \textbf{variation in sound}.
Here, the participants' conception of uncertainty involved high fluctuation, often described as a ``shaky'' or ``wavy'' quality, which also induced an ambiguous feeling in regards to the predicted data.
This is reflected through the prevalent use of oscillation effects, with the use of the modulation sliders.
Several participants found this modulation to be an intuitive way for conveying the uncertainty, describing the resulting sonification as reflecting the variable temperature ranges of the predicted data.
Modulation was also not just an indicator of uncertainty, but was also often mapped to the uncertainty itself, with greater uncertainty corresponding to greater oscillation effects. 
While not as conspicuous, \textbf{sound embellishments}, such as white noise, were also found to be useful in describing uncertainty.
White noise created a ``fog of uncertainty'' (P08) distorting the underlying sound, which enhanced the perception of uncertainty.
However, multiple participants mentioned how an increase in uncertainty would not lead to an increase in white noise.
This could be connected to the \textbf{pleasantness of sound}, which participants tried to achieve.
This solidifies the idea that white noise could be used as indicator of uncertain data, while not being directly mapped to it.
Instrument choices were varied, with the electric synth and and the piano being the most frequently selected, while the horn and the trumpet were also utilized.

\subsubsection*{Ominous}
The \textbf{ominous theme} relates uncertainty to some sort of personal association with \textbf{eerie} \textbf{ecological sounds}, such as from horror movies.
Instrument choices varied, but compared to other themes, participants actively based their choices on personal experiences instead of using unknown sounds.
Participants attempted to reproduce sounds that matched their experiences, such as recreating stepping sounds that grew progressively closer.
The personal experience regarding this ominous feeling differed between participants, which led to slightly more varied sonifications compared to other themes.
Sonifications would use a very low pitch, with the volume being louder compared to the other themes.
Most used either white noise or modulation, sometimes neither, and seldom both.
Due to these differences in chosen auditory parameters but similarity of underlying explanations of participants, it appears that not just one singular auditory parameter induced this ominous or eerie feeling.
Interestingly, six of the ten ominous sonifications were produced by the eight female participants.
It could be argued that sound design choices directly used in horror movies could prove useful in this regard. 

\section{Summary and Discussion}
Through a co-design user study, we investigated how affective qualities of uncertainty and neutrality in visualizations can be conveyed using sonification. Participants designed sonifications for a visualization of historical (neutral) and predicted (uncertain) temperatures while explaining their reasoning. 
Their designs, which are available as supplemental materials \cite{DARUS}, employed a wide range of auditory parameters, informed by diverse considerations, including emotional associations and personal experiences. Rather than offering validated sonification mappings, we present hypotheses-generating results regarding sonification-affect mappings for \mbox{uncertainty visualization.} 

\subsection{Insights for Neutral Sonifications}
The resulting sonifications for neutrality can be categorized into two themes: the \textbf{clear theme} and the \textbf{relaxing theme}.

\paragraph{Clear theme:}
Sonifications were kept simple, with \textbf{minimal sound features} as participants aimed to convey the data in a clear and neutral manner.
Participants used instrument options they did not associate with musical instruments and other auditory qualities with a \textbf{lack of emotion}.
Clarity also included adequate loudness and \textbf{temporal aspects} that slowed down the sonification enough to ensure \textbf{perceptibility} of the data, which allowed users to follow the sonifications with the visualization.
Some participants aimed to use pitch to \textbf{map to data} or other aspects in the visualization. However, others pointed out that the variability in the underlying data would produce a sonification with too much fluctuation to be perceived as neutral. As a result, they chose sonification qualities that resulted in a more \textbf{stable} sound. 

\paragraph{Relaxing theme:}
However, other participants associated neutrality with relaxing and comforting qualities. Their sonifications included \textbf{sound embellishments}, such as white noise, associated with \textbf{ecological sounds}  perceived as calming, such as~rain.

\subsection{Insights for Sonifications that Feel Uncertain}
The sonifications for uncertainty and the reasoning behind them can be divided into two themes: the \textbf{wavy theme} and the \textbf{ominous theme}.
While there is overlap between auditory parameters used in both themes, the reasoning behind the choices differ in conceptual focus, functional objectives, and the intended interpretive cues for~listeners.

\paragraph{Wavy theme:}
Participants considered \textbf{variation} and \textbf{fluctuation in sound} suitable auditory qualities for conveying uncertainty.
This included oscillatory effects often achieved using the amplitude modulation or modulating the frequency of the amplitude modulation. 
However, some participants used other auditory parameters such as pitch to create a wavy quality in their sonifications.
Oscillation was also considered an appropriate mapping for changes in uncertainty. Participants increased either one or both amplitude modulation and frequency of amplitude modulation for future years, where the uncertainty increases.
One participant, P13, took \textbf{variation in sound} a step further and wanted uncertainty to represent true randomness of sound.
They wanted to \textbf{map to data} by mapping randomly generated pitches within the upper and lower bounds of the uncertainty bands.

\paragraph{Ominous theme:}
Participants related auditory qualities that associated uncertainty with a \textbf{strange}, scary, or \textbf{eerie} sentiment for their uncertainty sonifications. These auditory qualities were often related to \textbf{ecological sounds} such as horror movies and sounds they could not associate with instruments.
Eeriness in uncertainty sonifications was often created with deep, low pitches.
However, others considered the \textbf{alarming} aspects of uncertainty and conveyed that with a high pitched sound.

\subsection{General Sonification Insights}
Participants used or avoided certain auditory parameters regardless of whether they were designing for neutrality or uncertainty. 
In contrast, some auditory parameters were employed in both types of sonifications but were justified differently, often attributing diverging associations and emotions to the same auditory parameters. 
Therefore, these auditory parameters could be considered ambiguous in how they convey emotions and data.

For both sonifications, participants considered \textbf{pleasantness} an important attribute. Unpleasant auditory qualities such as too much white noise, extreme amounts of modulation, too much volume, and certain instruments were avoided.
Some auditory parameters, including amplitude modulation and frequency of amplitude modulation were considered \textbf{goofy} and therefore found unsuitable for uncertainty or neutrality. Such auditory parameters may be better suited when sonifying a positive, comedic affect.
Instrument options that participants associated with recognizable musical \textbf{instruments} were generally perceived as neither neutral nor uncertain. 
Consequently, instrument options that are unnatural in the sense that they do not readily map onto familiar musical instruments may be better suited for representing these affective qualities.

\textbf{White noise} was considered by some participants to be neutral because of its \textbf{relaxing} properties.
This is not an unusual sentiment as white noise is is known to help some to relax and fall asleep~\cite{RIEDY2021101385}.
On the other hand, white noise was also a popular choice for the uncertain sonifications in the \textbf{ominous theme} and \textbf{wavy theme}.
Participants considered white noise to be \textbf{eerie}.
Because white noise was considered suitable for conveying both neutral and uncertain affect, we suggest avoiding it, as its ambiguous interpretation might lead to unintended understandings of the data.

\subsection{Semiotics or Affective Associations}
Affective visualization and semiotics largely overlap as both examine how visualization design mappings relate to user interpretations.
Semiotics investigates how users relate stimuli to (data) concepts \cite{ballatore2019sonifying, bertin1983semiology, maceachren2012visual, vriend2025semiotics}, whereas affective visualization examines the emotional components of (visual) stimuli \cite{lan2024affective, Lee2023Affective}.

From our results, we noticed that participants associated neutrality and uncertainty to emotional components and abstract concepts, often interchangeably.
Neutrality was associated with affective qualities such as \textbf{boring} or \textbf{lack of emotions} but also to \textbf{abstract qualities}, concepts not related to emotions such as predictability.
Uncertainty was associated with affective qualities such as \textbf{ominous} and \textbf{eerie} 
but also to \textbf{ecological sounds} such as video games, and to \textbf{abstract qualities} like \textbf{strange} or \textbf{unusual}. 

This begs the question if there is a difference between semiotics and affective visualization.
Our results suggests that semiotics and affective visualization are related. However, we do not know whether participants related auditory qualities to semiotic associations that had an affective component, or whether participants associated auditory qualities to affect that 
led to semiotic associations. 
Another potential source of associations are the innate reactions we have to certain kinds of sounds. Juslin et al.~\cite{juslin_emotional_2008} highlight that certain auditory qualities (such as loudness and fluctuations) were important to be attentive to for our survival, and might therefore evoke related types of emotions.

\section{Limitations and Future Work}
One participant, P05, was interrupted during their study session. 
This could have had some negative effect on the participant such as disturbing their current train of thought, which could lead to forgetting their reasoning for their auditory parameter choices. 
Such unexpected interactions can also induce some emotions, such as confusion.
However, from the transcripts, it did not seem this interruption seriously impacted the participant.
Nonetheless, we coded the interruption for transparency sake, which is reflected \mbox{in \autoref{fig:Frequency_Codes_Unique-1}.}

The context and meaning of the data may have shaped sonification designs, at least to some extent.
The increasing of predicted future temperatures can be associated with global warming, which for many has a negative connotation. 
This negative connotation may have influenced sonification designs and decisions of some participants.
For instance, P20 explicitly based their uncertainty sonification on the negative implications associated with climate change, explaining:
``\textit{So it technically gets bad because like, I feel like the air temperature, if it's like getting up over the years, it's not good because of climate change. So I would say that it's like, if it's uncertain, but it also should convey some negativity to some extent. Especially because [...] it's getting bad and it's getting higher.}''
Hence, it is important to interpreted our results and their implications in the context of the data used in the study.
Future research should investigate sonification for other uncertainty visualizations and their data types and understand whether participants are primarily sonifying the visual encoding, the uncertainty itself, the semantics of the data, or the broader real‑world context it represents. Such research should include datasets with a variety of context such as a dataset with a hopeful uncertainty context.  

The qualitative nature of this work and the small and relatively homogeneous participant pool mean that our findings offer limited generalizability. 
As a result of this exploratory approach, this work provides hypothesis-generating, rather than conclusive evidence.
Future work could expand upon our study by formulating and quantitatively testing hypotheses based on our findings with a broader and more diverse group of participants.
For example, a future study could validate our findings by involving a separate group of listeners who assess whether the resulting sonifications effectively convey uncertainty, neutrality, or the underlying data. 
Another avenue for future work could examine whether the themes identified during the co-design process emerge from auditory perception or other \mbox{cognitive processes.}

This work focused on the affective components of neutrality and uncertainty. However, some participants mentioned certainty instead of neutrality. Future research could examine certainty, for example, by identifying auditory qualities that elicit a sense of confidence, and how these responses compare with the affect of neutrality and uncertainty.

We analyzed our data using RTA, a qualitative method of analysis rooted in an interpretivist paradigm that stands in stark contrast with quantitative approaches.
We chose RTA because it is a good fit for our research as we investigated not only what auditory parameters are suitable for sonifying the affective aspects of uncertainty and neutrality but also why.
We focused on understanding how users experience and make sense of auditory parameters and their affective aspects.
RTA allowed us to identify patterns in that data, which yielded rich, nuanced, and insightful findings.
We ensured a rigorous analysis process, in accordance with RTA guidelines~\cite{Braun2024RTA}, by drawing on diverse expertise within our research team, which included a researcher with an HCI background, a computer scientist, a sonification expert, and a visualization expert. Throughout the study, we documented our research practice and decisions as transparently as possible. As discussed above, this qualitative approach could be complemented by future quantitative studies for \mbox    {hypothesis testing.}

\section{Conclusion}
We investigated how sonification can be employed to convey the affective component of uncertainty and neutrality in a visualization of temperature predictions by asking users to design sonifications that feel uncertain or neutral. 
In a co-design user study, participants manipulated auditory channels to express a feeling of uncertainty and neutrality in a visualization of past and future temperatures.
The resulting sonification artifacts and the reasoning participants articulated while creating the sonifications were analyzed to offer hypotheses-generating insights for sonifications that can be integrated with uncertainty visualization to convey the affect of uncertainty. 
Both the sonifications created by participants and the codebook resulting from the analysis are provided as \mbox{supplemental materials~\cite{DARUS}.} 

\section*{AI Use Statement}
We used an internal AI tool provided by our university and Microsoft´s Copilot for text editing. Perplexity was used for brainstorming limitations regarding RTA.
The internal AI tool was also used for finding a reference investigating the calming properties of white noise \cite{RIEDY2021101385}.

\acknowledgments{
The authors thank the participants for their time and effort. This work was funded by the 
Deutsche Forschungsgemeinschaft (DFG, German Research Foundation)---Project-ID 251654672---TRR~161.
}

\bibliographystyle{abbrv-doi-hyperref-narrow}

\bibliography{revised_ref}
\end{document}